\documentclass[12pt,preprint]{aastex}         
\usepackage{graphicx}
\usepackage{epstopdf}
\usepackage{amsmath}
\def\be{\begin{equation}}
\def\ee{\end{equation}}

\def\lsim{\lower 2pt \hbox{$\, \buildrel {\scriptstyle <}\over
         {\scriptstyle \sim}\,$}}

\begin{document}
\newcommand{\figureout}[3]{\psfig{figure=#1,width=5.5in,angle=#2} 
   \figcaption{#3} }

\title{Multi-TeV Emission From the Vela Pulsar}

\author{Alice K. Harding\altaffilmark{1}, Constantinos Kalapotharakos\altaffilmark{1}$^,$\altaffilmark{2}, Monica Barnard\altaffilmark{3} and Christo Venter\altaffilmark{3}}
\altaffiltext{1}{Astrophysics Science Division,      
NASA Goddard Space Flight Center, Greenbelt, MD 20771}
\altaffiltext{2}{University of Maryland, College Park (UMDCP/CRESST), College Park, MD 20742}
\altaffiltext{3}{North-West University, Private Bag X6001, Potchefstroom 2520, South Africa}
 

\begin{abstract}
Pulsed emission from the Vela pulsar at energies above 3 TeV has recently been detected by the H.E.S.S. II air-Cherenkov telescope.  We present a model for the broad-band spectrum of Vela from infra-red (IR) to beyond 10 TeV.  Recent simulations of the global pulsar magnetosphere have shown that most of the particle acceleration occurs in the equatorial current sheet outside the light cylinder and that the magnetic field structure is nearly force-free for younger pulsars.  We adopt this picture to compute the radiation from both electron-positron pairs produced in polar cap cascades and from primary particles accelerated in the separatrix and current sheet.  The synchrotron spectrum from pairs resonantly absorbing radio photons at relatively low altitude can account for the observed IR-optical emission.     We set the parallel electric field in the current sheet to produce the {\it Fermi} GeV emission through curvature radiation, producing particles with energies of 30-60 TeV.  These particles then produce Very-High-Energy emission up to around 30 TeV through inverse-Compton scattering of the IR-Optical emission.  We present model spectra and light curves that can match the IR-Optical, GeV and make predictions for the multi-TeV emission.
\end{abstract} 


\pagebreak
  
\section{Introduction}

The study of high-energy emission from rotation-powered pulsars entered a new era ten years ago with the detection of the Crab pulsar at energies above 25 GeV by the MAGIC ground-based air-Cherenkov telescope (Aliu et al. 2008).  This was quickly followed by pulsed detections above 100 GeV by VERITAS (Aliu et al. 2011), up to 400 GeV (Aleksic et al. 2012) and up to 1.5 TeV  with MAGIC (Ansoldi et al. 2016).  Most recently, H.E.S.S. Collaboration has announced detection of pulsed emission from the Vela pulsar from 20 to 100 GeV (Adballa et al. 2018)  and also, remarkably, above 3 TeV (Djannati-Atai et al. 2017).  This result by itself constrains the energy of the accelerated particles in pulsar magnetospheres to at least a few TeV.  Finally, the MAGIC telescope has announced detection of pulsed emission from the Geminga pulsar (Lopez et al. 2018), the first Very-High-Energy (VHE) emission from a middle-aged pulsar.  For all three pulsars, detected emission up to 1 TeV appears to connect to the GeV spectra measured by the {\it Fermi Gamma-Ray Space Telescope}, but the emission from Vela above 3 TeV may also be a separate component.

A number of emission models ascribe the {\it Fermi} spectrum measured from 100 MeV up to 30-50 GeV to curvature radiation (CR) (e.g. Romani 1996, Hirotani 2001, Harding et al. 2008, Kalapotharakos et al. 2017, 2018).  The presence of pulsed emission above 100 GeV challenges this CR picture as the sole high-energy radiation mechanism and requires at least one additional emission component.  Inverse-Compton (ICS) radiation has been suggested as a mechanism to produce the VHE emission from the Crab pulsar (Lyutikov et al. 2012, Du et al. 2012, Harding \& Kalapotharakos 2015 [HK15]).   Lyutikov (2013) modeled the Crab VHE emission as cyclotron-self Compton emission from electron-positron pairs, where pairs from an outer gap scatter their own synchrotron radiation (SR).   HK15 modeled the Crab optical to TeV emission with a synchrotron-self Compton (SSC) model, where pairs from the polar cap scatter their own SR in the outer magnetosphere.  Other models have proposed that synchrotron emission from particles accelerated by reconnection in the current sheet could reach up to TeV energies through Doppler boosting (Uzdensky \& Spitkovsky 2014, Iwona \& Petri 2015).  

The emission from Vela above 3 TeV cannot be radiated by secondary pairs since their spectra do not extend beyond 1 TeV, even for the case of the Crab pulsar.  This emission is therefore likely to be radiation from the accelerated particles that are thought to reach energies of at least 10 TeV.  Rudak \& Dyks (2017) [RD17] modeled this VHE emission from Vela using an outer gap model extending along the last open field lines from the null charge surface ${\bf \Omega \cdot B} = 0$ to near the light cylinder (LC), $R_{\rm LC}  = c/\Omega$, where $\Omega$ is the pulsar rotation rate.   Primary particles accelerated in the gap scatter the observed infra-red (IR) to optical emission, placed along the gap inner edge and assumed to come from SR of pairs produced in the gap.  Although the pair SR was not modeled they showed that, normalizing the IR-optical component to that observed, primaries whose CR spectrum matches the {\it Fermi} spectrum produce a significant component of ICS emission extending up to around 10 TeV.  They note that the IC scattering of the lower part of the IR-optical spectrum occurs in the Thompson limit, while scattering the upper end of the IR-optical spectrum will reach the Klein-Nishina limit so that the IC spectral energy distribution (SED) peaks and then cuts off at the particle energy.  The observed emission around a few TeV from the Vela pulsar thus provides a lower limit to the particle acceleration energy in pulsars.

In this paper we model the broad-band spectrum of the Vela pulsar from IR to beyond 10 TeV, using the most up-to-date physical description of pulsar magnetospheres.  Recent MHD (Kalapotharakos et al. 2014, 2017) and Particle-in-Cell (PIC) simulations (Cerutti et al. 2016, Philippov \& Spitkovsky 2018, Brambilla et al. 2018, Kalapotharakos et al. 2018) of the global pulsar magnetosphere have shown that most of the particle acceleration occurs in the equatorial current sheet outside the LC rather than in outer gaps inside the LC.  In young and middle-aged pulsars electron-positron pairs are produced in  cascades, creating a near-force-free magnetosphere where electric fields parallel to the magnetic field are almost fully screened.  We have adopted this picture, using a force-free magnetic field structure to model radiation from pairs and accelerating particles that together can produce a broad spectrum of radiation.  

\section{Model and Assumptions}

The computations performed in this paper use an expanded and updated version of the model presented in HK15.  The first major improvement is an expanded spectral energy range for the radiation, computing the emission over 18 decades from IR ($10^{-3}$ eV) to VHE (100 TeV) energies, whereas HK15 computed spectra from only 3 eV to 1 TeV.  This will turn out to be of crucial importance both for including more of the soft photons whose scattering is in the Thompson limit, and for modeling the scattered emission at the highest energies.
The second update is a more accurate computation of the particle trajectories and their radii of curvature which gives a more accurate determination of the energy of the accelerating particles and of their spectrum.  A full description of the new trajectory calculation will appear in a forthcoming paper (Barnard et al. 2018).  
The third update to the calculation is to set the parallel electric field ($E_\parallel$) of accelerating particles to a lower value inside the LC rather than set a constant high value of $E_\parallel$ from the neutron star surface to $2 R_{\rm LC}$.  This change provides better agreement with recent global MHD and PIC pulsar models that have shown that the particle acceleration takes place primarily near the current sheet outside the LC.  The fourth update is to inject the electron-positron pairs only above the PC in regions where the global force-free current density enables pair cascades (details are described below).  The fifth improvement is to model the radiation from accelerating particles inside and outside the LC with synchro-curvature (SC) radiation, although their radiation at GeV energies is mostly CR.  In HK15, both the accelerating primary particles (as well as the pairs) could acquire pitch angles at low altitude through cyclotron resonant absorption of radio photons, but their SR and CR were treated separately.

\begin{figure} 
\includegraphics[width=150mm]{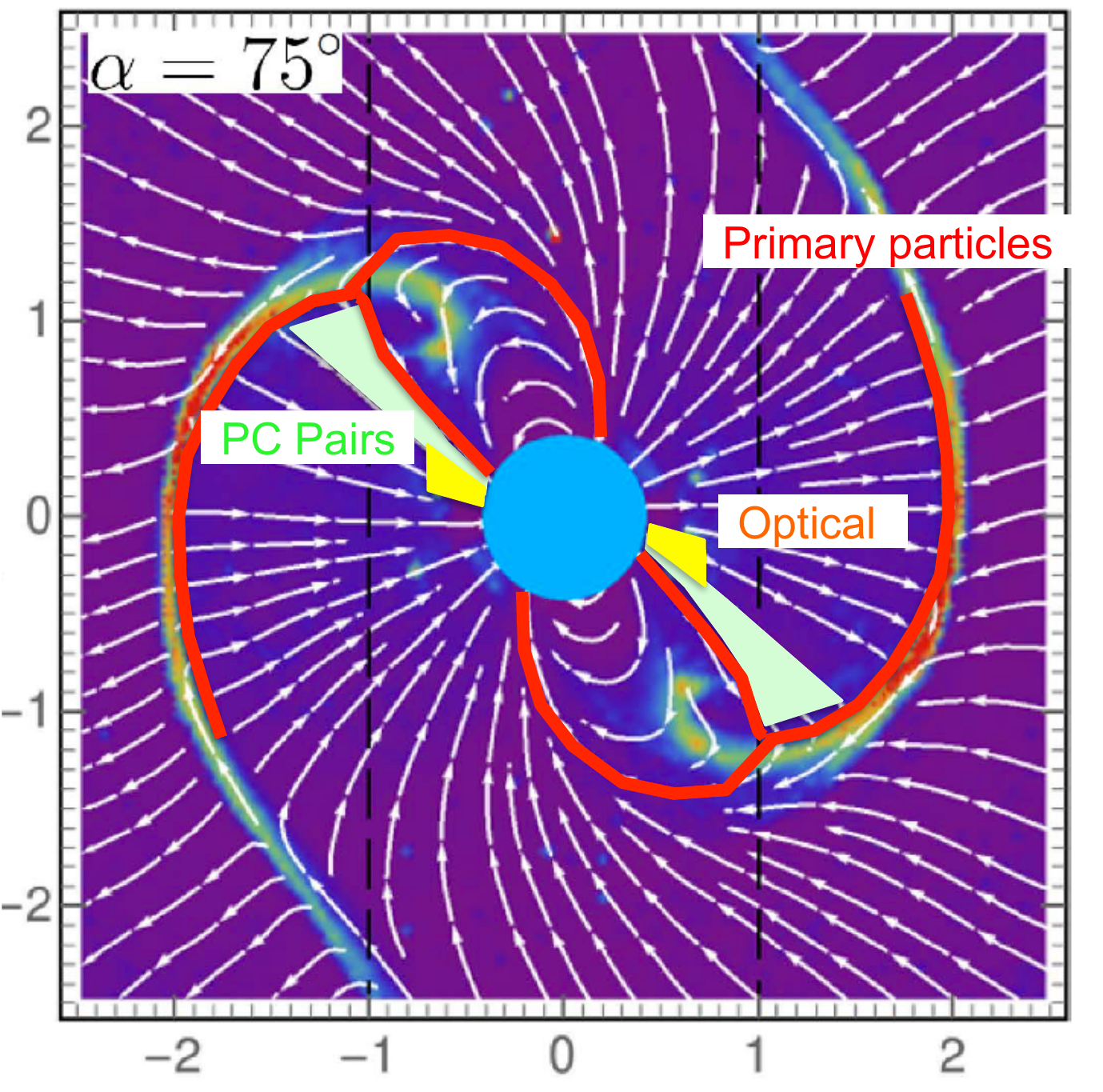}
\caption{Schematic illustration of model components.  White lines are magnetic field lines projected on the magnetic-spin axis plane of a near-force-free magnetosphere for inclination angle $\zeta = 75^\circ$ (from Kalapotharakos et al. (2018)).  Red lines represent the trajectories of accelerated particles injected at the NS surface and extending into the current sheet.  Green regions represent the PC pairs, injected at the NS surface and emitting SR optical/IR radiation (yellow regions) at low altitude where they can resonantly absorb radio photons.}  
\label{fig:magneto}
\end{figure}

As in HK15, we use the magnetic and electric field structure of a global force-free (${\bf {E \cdot B} = 0}$) magnetosphere for our radiation modeling, which provides a fairly accurate description of the field of an energetic young pulsar.  The fields at each point are interpolated from a numerically-computed Cartesian grid centered on the neutron star and cubic computational volume of size $(4.0 R_{\rm LC})^2$.  We define open volume coordinates (Dyks et al. 2004) that cover the polar caps (PC) in concentric rings labeled by $r_{\rm ovc}$, with $r_{\rm ovc} = 0$ at the magnetic pole and $r_{\rm ovc} = 1$ at the PC rim.  Each ring is divided into 360 azimuthal segments.  We inject two populations of particles at the neutron star surface in prescribed regions on the PC.  ``Primary" particles (electrons or positrons) are injected between $r_{\rm ovc} = 0.90$ and 0.96 with low Lorentz factor ($\gamma = 200$) and a spectrum of electron-positron pairs  are injected between $r_{\rm ovc} = 0.80$ and 0.90.   From both recent global PIC simulations (Brambilla et al. 2018) and studies of PC pair cascades (Timokhin \& Arons 2013), it seems possible that pair cascades near the neutron star surface supply the particles that reach the current sheet and are accelerated to high energy there, as well as the lower-energy pairs that are responsible for screening the parallel electric fields inside the LC.  So in reality, both the ``primary" particles that are injected on field lines near the PC rim that connect to the current sheet and are accelerated there, and pairs originate from PC pair cascades but on different field lines (see Figure \ref{fig:magneto}).  We treat them separately in this model.  The injected pair spectrum used here is from a separate local calculation of a  pair cascade for the Vela pulsar (see HK15, Fig. 1) with a pair multiplicity of $6 \times 10^3$.  Simulations of PC pair cascades (Timokhin \& Arons 2013) that are compatible with global current density $J$ follow a distinct pattern; pair cascades operate only in regions of the open field where $J/J_{\rm GJ} > 1$ or  $J/J_{\rm GJ} < 0$, where $J_{\rm GJ} = \rho_{\rm GJ} c$ and $ \rho_{\rm GJ} = {\bf -\Omega \cdot B}/2\pi$ is the local Goldreich-Julian charge density (Belodorodov 2008).  The return (anti-GJ) current regions $J/J_{\rm GJ} < 0$ connect to the current sheet and are bounded by the super-GJ regions.  We therefore inject the pairs in the region of anti-GJ current, which for the magnetic inclination $\alpha = 75^\circ$ we adopt for Vela is roughly the azimuthal regions on the lower half of the PC opposite to the rotation axis and towards the current sheet (Timokhin \& Arons 2013).  

The particle trajectories are computed, as described in HK15, by determining the velocity as the sum of a drift component and a component parallel to the local magnetic field, where the fields in this case are those of the force-free magnetosphere.  However, as noted above, we do not compute the trajectories as the particles are radiating and accelerating as in HK15, but we now pre-compute trajectories and radii of curvature for each injected particle initial position on the PC, assumed to be energy-independent, and store them for later interpolation of positions in the particle dynamics and radiation calculation.  For the dynamics of the ``primary" particles we assume different values of constant parallel electric field $E_\parallel$, with a lower value inside the LC, $E_\parallel^{\rm low}$ and higher value outside the LC, $E_\parallel^{\rm high}$.  
Kalapotharakos et al. (2018, Fig. 7 and 9) presented spatial distributions of particle energy and $E_\parallel$ in magnetospheres with different pair injection rates, showing that for the highest injection rates, applicable to young pulsars, the regions of highest particle energy and $E_\parallel$ occur in the current sheet outside the LC, with lower particle energy and $E_\parallel$ along the separatrix below the LC.
All particles can radiate a mix of SC, SSC and ICS and the detailed treatment of all of these processes as well as the calculation of their emission directions can be found in HK15 and Harding et al. (2008, H08).  For the SC emission and energy loss rate we use the formulae given by Torres (2018).  Both primaries and pairs produce SC in the SR regime through resonant cyclotron absorption of radio photons (Lyubarski \& Petrova 1998), which maintains moderate pitch angles to balance synchrotron losses (H08).  The radio emission is radiated at an assumed altitude of $r_{\rm radio}$.  Since the pairs are not accelerating, their Lorentz factors are quite low ($\gamma < 10^5$) and their SC is entirely in the SR regime.  The calculation of the radiation takes place in two phases, the first being the SC of pairs and primaries to store the emissivities and the second being scattering of that radiation.  To model the Vela pulsar multi-TeV emission in this paper, we consider two separate calculations.  In the first, we inject all particles as described above but also include an additional ``toy" model of the observed IR-optical spectrum of Vela, with a flux level and power law index to match the measured spectrum of Shibanov et al. (2003).  To do this, a power-law spectrum from $E_{\rm min}$ to  4 eV,
\be
L_{IR} =  L_0  \left({E\over E_0}\right)^{-1},  
\ee
where $L_0 = 8 \times 10^{40}\,\rm ph\,s^{-1}\,eV^{-1}$ and $E_0 = .04$ eV, is radiated uniformly along the trajectories of the pairs from the NS surface $r_{\rm min, IR} = R_{\rm NS}$ to a radius  to $r_{\rm max, IR} = 0.5 R_{\rm LC}$.   We will allow the spectrum to extend below the observed IR-optical spectrum to a lower limit as small as $E_{\rm min} = .005$ eV, similar to the model of RD17.  The local emissivity of this component, $\epsilon_{\rm IR} = L_{\rm IR} / V_{\rm IR}$, is stored in an array in the first phase to be used for calculation of the photon density along particle trajectories in the scattering phase of the modeling, where 
\be
V_{\rm IR} =  \pi\int^{r_{\rm max}}_{r_{\rm min}} rdr\int^{\theta_{\rm out}}_{\theta_{\rm in}}d\theta\, r\theta(r) \simeq {\pi \over 2R_{\rm LC}}(r_{\rm max, IR}^4-r_{\rm min, IR}^4)(r^2_{\rm ovc,out}-r^2_{\rm ovc,in})
\ee
is the volume of the IR radiation, with $r_{\rm ovc,in} = 0.80$ and $r_{\rm ovc,out} = 0.90$.  The photon density for scattering by the particles is then computed according to Eqns (30) and (31) of HK15.  We used first the ``toy" model for the IR emission since we can easily adjust the parameters to fit the observed data.  In the other calculation, we replace the IR toy-model spectrum with the pair SR spectrum for the soft scattered radiation.  Both the IR and the pair SR radiation are restricted to the field lines of pair injection, as described above.  All of the radiation is accumulated in sky maps of observer angle  and phase (rotational colatitude and azimuth) for 54 energy bins, from which we create spectra and light curves.  The parallelized computations were carried out on the Discover cluster of the NASA Center for Climate Simulation (NCCS).

\section{Results}

\begin{figure} 
\includegraphics[width=180mm]{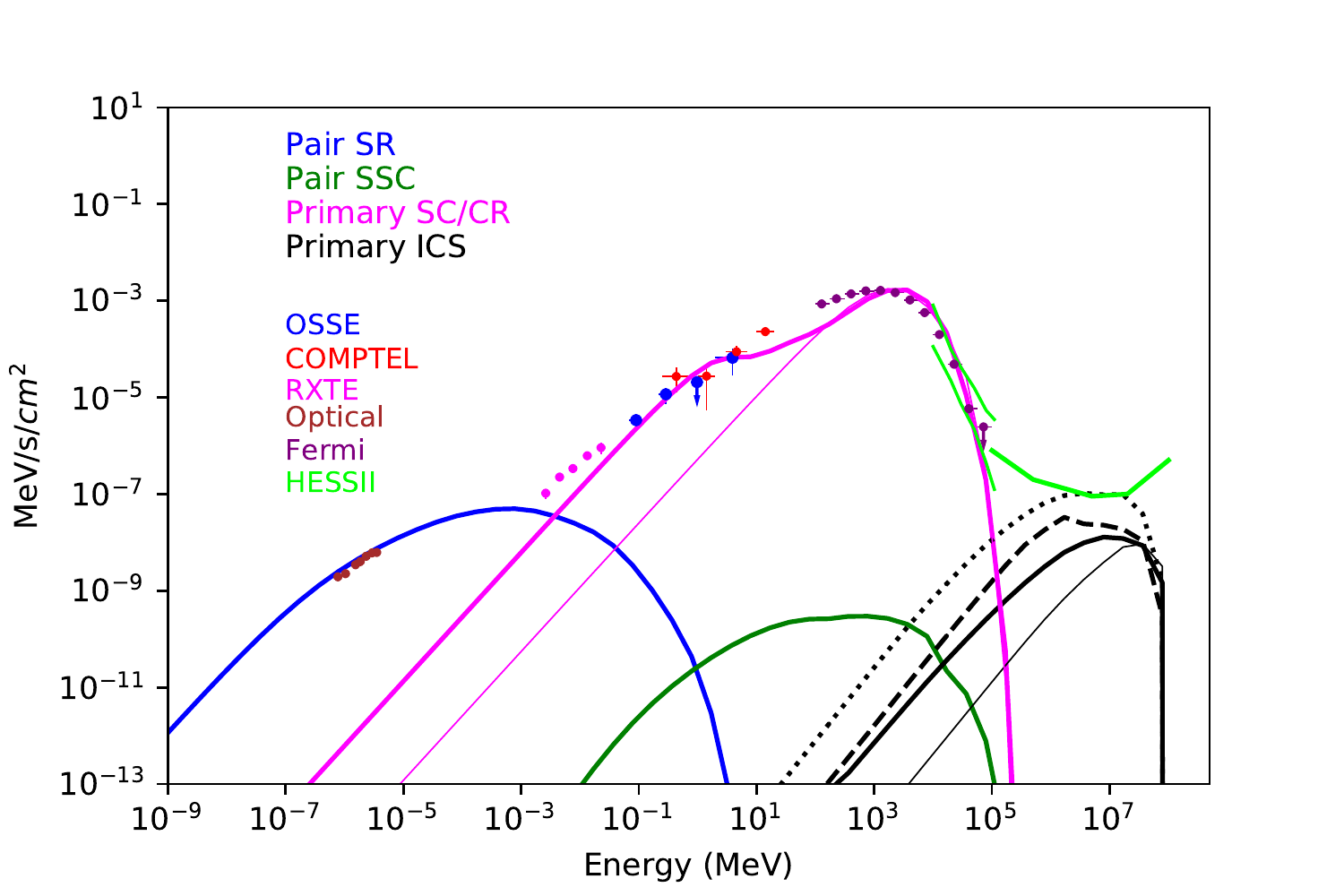}
\caption{Model spectral energy distribution of phase-averaged emission from accelerated particles and pairs (as labeled), for magnetic inclination angle $\alpha = 75^\circ$ and viewing angle $\zeta = 65^\circ$.  The solid black lines are the ICS components from accelerated SC-emitting (thick line) or CR-emitting (thin line) primaries scattering the pair SR component (blue solid line), while the dashed  and dotted black lines are the spectra from SC-emitting primaries scattering toy model soft IR/optical photons with energy range (.5 - 4 eV) and (.005 - 4 eV) respectively.  Data points for the Vela pulsar are from Abdo et al. (2013; http://fermi.gsfc.nasa.gov/ssc/data/access/lat/2nd\_PSR\_catalog/), Shibanov (2003) and Harding et al. (2002).   The H.E.S.S. II detection (Abdalla et al. 2018) and high-energy sensitivity are also shown.}  
\label{fig:spec}
\end{figure}

Figure \ref{fig:spec} shows our modeled broad-band SED of Vela with observed data points for magnetic inclination angle $\alpha = 75^\circ$ and a viewing angle of $\zeta = 65^\circ$.  We find that the pair SR matches very well the observed optical spectrum in both magnitude and power law index over that energy band.  Based on this match, we postulate that the Vela IR-optical emission is SR radiated at relatively low altitude by pairs from the PCs.  This pair SR component was also shown in Fig. 6 of HK15, but its spectrum was not computed below 3 eV.   The SC of primaries was adjusted to match the peak of the {\it Fermi} spectrum by setting the constant parallel electric field value in the current sheet, $eE_\parallel^{\rm high}/mc^2 = 0.2$.  
Such a value, equivalent to $E_\parallel \sim .01\,B_{\rm LC}$ where $B_{\rm LC}$ is the magnetic field strength at the LC, is consistent with the $E_\parallel$ values of PIC simulations, scaled to the Vela pulsar parameters and pair injection rate (Kalapotharakos et al. 2018).
This value of $E_{\parallel}$ will determine the maximum energy of the ICS component at multi-TeV energy.  We also assume a particle multiplicity of $10J_{\rm GJ}$ to match the flux level of the {\it Fermi} GeV spectrum which will determine the flux level of the ICS emission.  This multiplicity of accelerated particles near the current sheet is in agreement with what is needed in global simulations to account for the $\gamma$-ray luminosity of Vela (Kalapotharakos et al. 2017, 2018).  The value of $E_\parallel$ below the LC was set to $eE_\parallel^{\rm low}/mc^2 = 0.04$.  With this $E_\parallel$ distribution the radiation is primarily SR at low altitude becoming predominantly CR outside the LC.  The pure CR spectrum of particles accelerating only in the current sheet is also shown in Figure \ref{fig:spec} (thin magenta line).

These primaries then scatter soft photons from either the IR toy-model component or the pair SR to produce the VHE component extended up to around 30 TeV.  The upper limit of this VHE spectrum occurs at the maximum energy of the primary particles and thus is an excellent diagnostic of the pulsar acceleration.  Figure  \ref{fig:spec}  shows the spectrum of scattering of both of the two optical/IR components by the SC-emitting particles.  Scattering of the toy IR spectrum is shown for two values of the lower limit: $E_{\rm min} = .005$ eV (black dotted line) and $E_{\rm min} = .5$ eV (black dashed line), while the scattered pair SR for pairs resonantly absorbing radio photon emission placed at $r_{\rm radio} = 0.2 R_{\rm LC}$ (black thick solid line).  The scattered pair SR spectrum is similar in shape and flux to the scattered toy IR spectrum only for the case of higher $E_{\rm min}$.  The scattered toy IR spectrum for the lower value of $E_{\rm min}$ has a much higher flux since there are more soft photons and the extrapolation of the power law significantly exceeds the flux of the pair SR spectrum.  If the pair SR is the correct representation of the optical/IR emission, then a scattered toy IR spectrum with the higher $E_{\rm min}$ is more realistic.  The peak fluxes of the scattered spectra are approaching or exceeding the H.E.S.S. II sensitivity and it is therefore possible that such a component will account for the observed multi-TeV emission.  We also show for comparison the ICS component of particles just emitting CR in the current sheet scattering the pair SR (thin black line).  It is evident that allowing low-level acceleration inside the LC enhances the lower energy part of the ICS component. The spectrum of scattering (SSC) from pairs is also shown (the green curve in Fig. \ref{fig:spec}) and is much lower in both energy and flux.  Both the components from primary and pair scattering were presented in Fig. 6 of HK15, with both being significantly higher in this calculation because of the much lower-energy extension of the spectrum of soft photons.  This extension is crucially important since the scattering of this emission by primaries is lowered by Klein-Nishina reductions above about 0.1 eV, but the scattering of lower-energy photons takes place in the Thompson limit.  This can increase the scattered flux by several orders of magnitude.  The SC emission from the primaries does not completely account for the hard X-ray/soft $\gamma$-ray emission observed by {\it RXTE, OSSE} and {\it COMPTEL}, but adding a more complex $E_\parallel$ distribution may give an improved match over our simplified model.

We also tried adding a spectral component to our toy IR model to include the recently discovered pulsed emission at 97.5 - 343.5 GHz from ALMA (Mignani et al. 2017).  This produced an ICS component that was at least one order of magnitude above H.E.S.S. sensitivity.  We therefore conclude that the sub-millimeter ALMA emission must be radiated at very low altitude and is more likely to be an extension of the radio emission.

\begin{figure} 
\includegraphics[width=90mm]{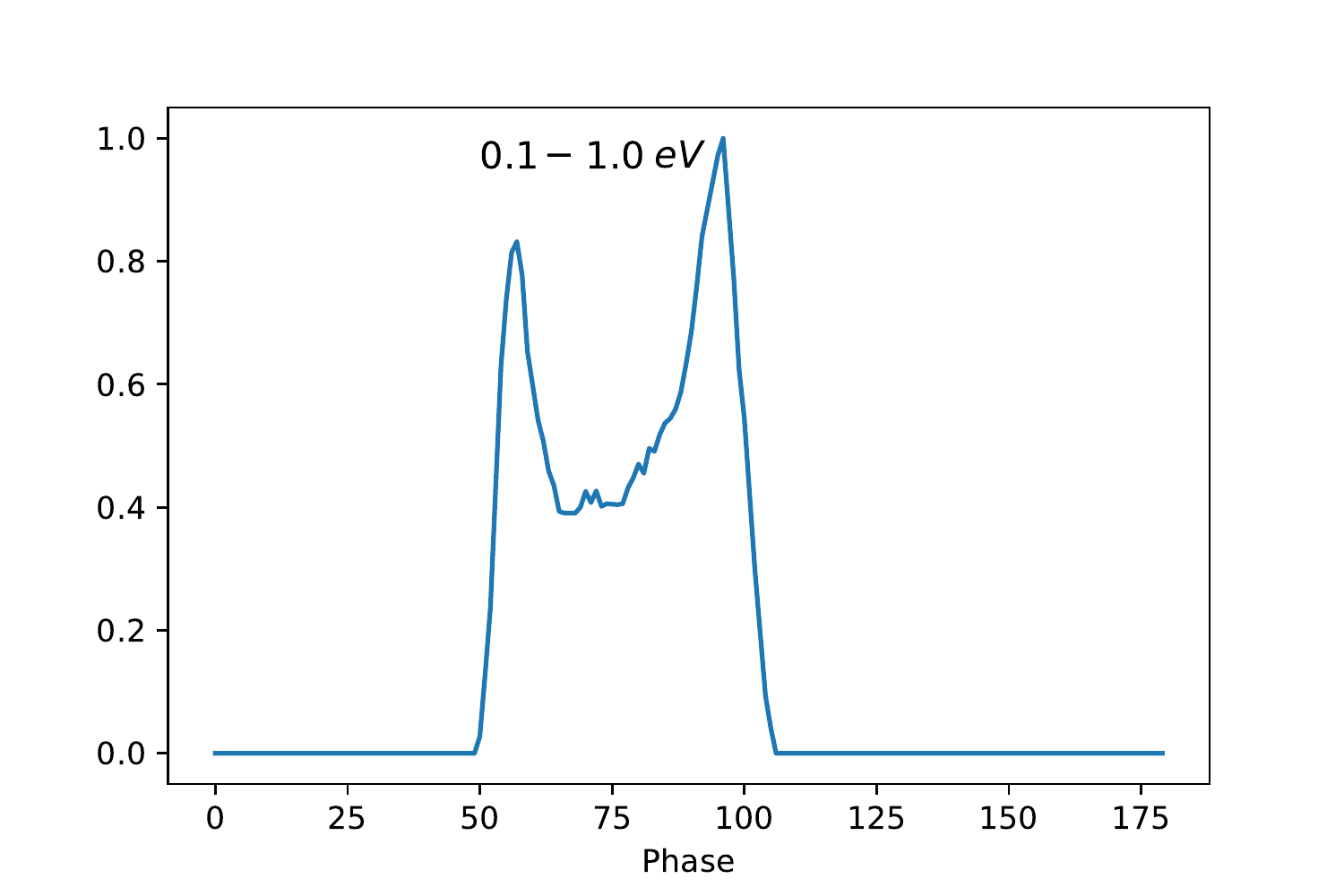}
\includegraphics[width=90mm]{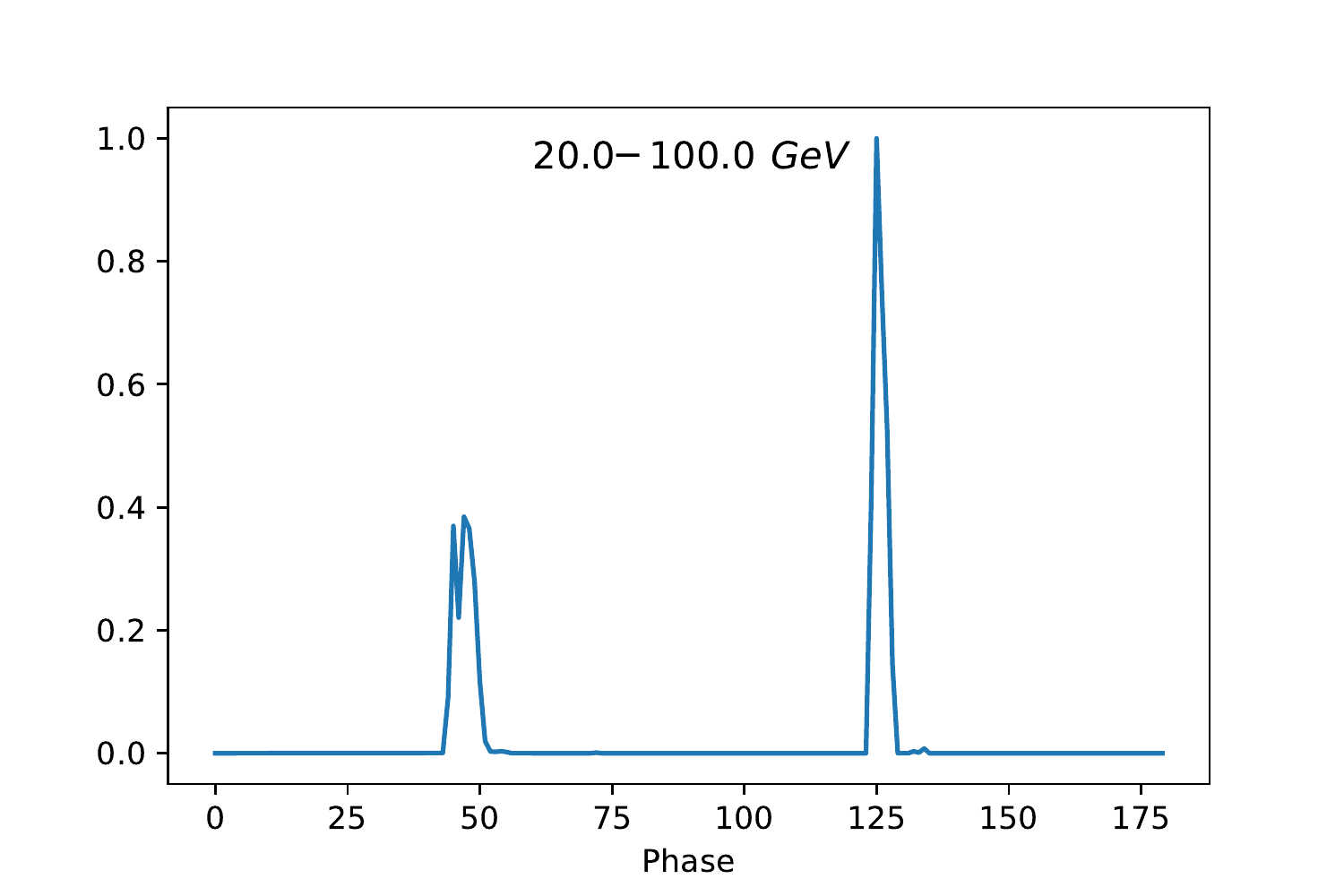}
\includegraphics[width=90mm]{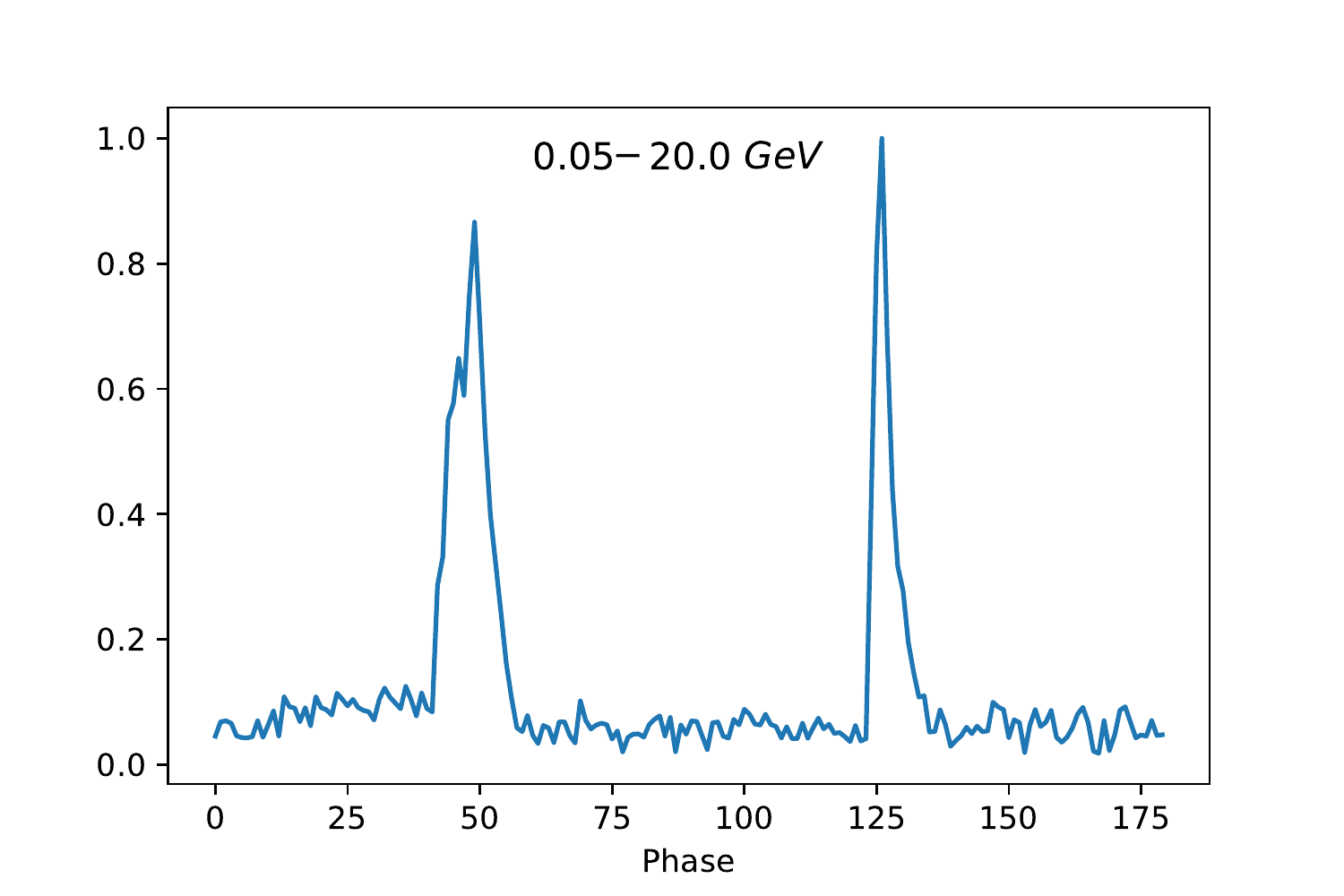}
\includegraphics[width=90mm]{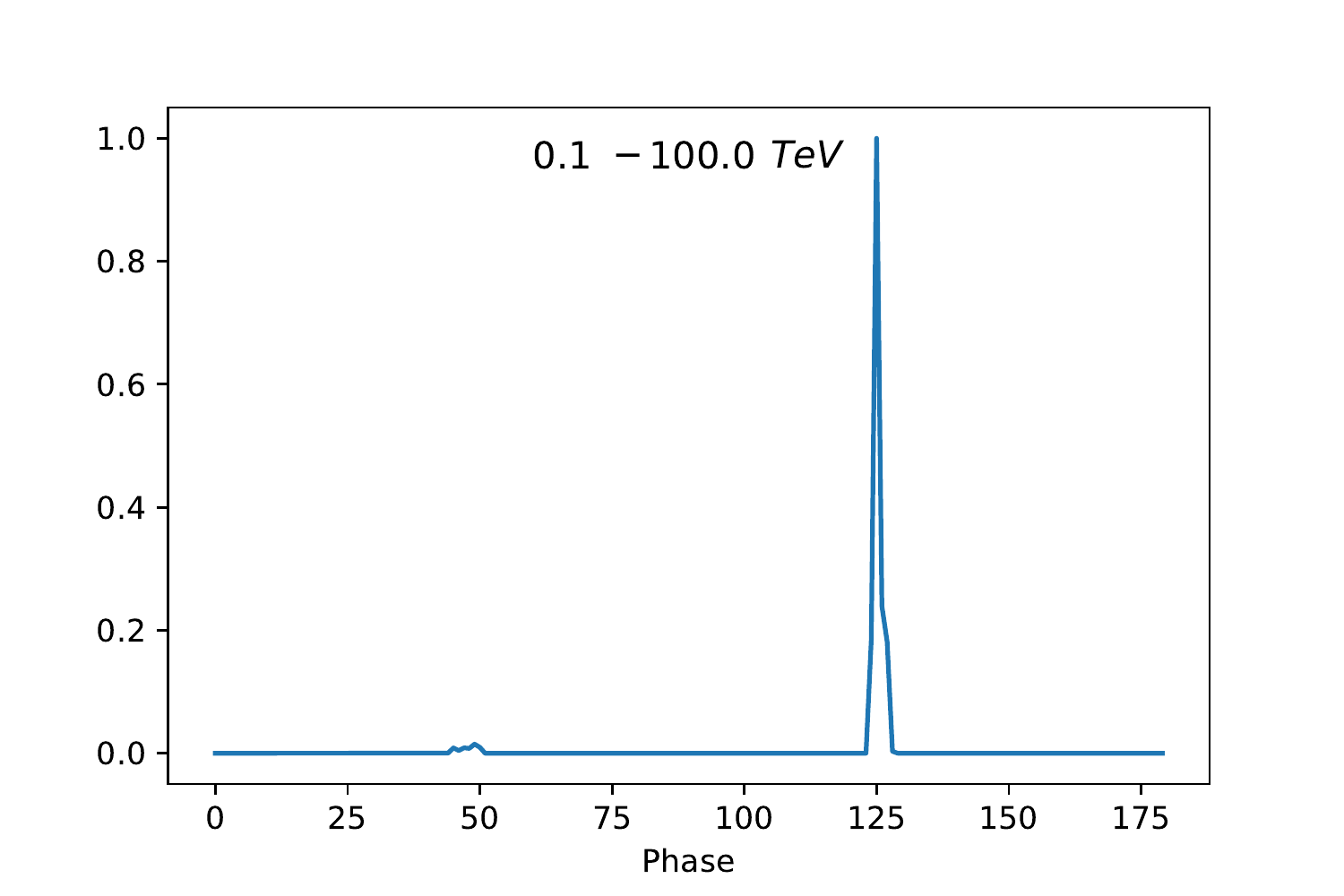}
\caption{Model light curves for emission in the IR/optical band (0.1 - 1 eV), and three different $\gamma$-ray energy bands, for magnetic inclination angle $\alpha = 75^\circ$ and viewing angle $\zeta = 65^\circ$.  Phases are in degrees.}
\label{fig:LCs}
\end{figure}

The model light curves are displayed in Figure \ref{fig:LCs} for the IR-optical, 0.05 - 20 GeV, 20 - 100 GeV and TeV bands.  The absolute phases of the peaks are not accurate in this calculation because we have used azimuthally independent distributions of $E_\parallel$ for the primaries.  The resulting phase of the first $\gamma$-ray peak (0.25) relative to the radio phase (around 0) is larger than observed for Vela (0.15).  It was shown in both global MHD dissipative models (Kalapotharakos et al. 2014) and PIC models (Kalapotharakos et al. 2018), where the non-azimuthally symmetric $E_\parallel$ distribution is determined self-consistently, that the phase lag with the radio peak is smaller and matches that observed.  The $\gamma$-ray peak separation at GeV energies is however similar to what is observed.  Our model light curves in the 0.05 - 20 GeV and 20 - 100 GeV bands reproduce both the observed narrowing of the peaks and the relative decrease of the flux of the first peak.  The light curve of the primary ICS consists primarily of a single peak at the phase of the second $\gamma$-ray peak.  This makes sense since the highest-energy photons from the highest-energy particles make up the second peak of the {\it Fermi} light curve and that peak persists up to the highest energies of the SC spectrum while the first peak decreases in relative flux.   Since in this model the $E_\parallel$ distribution is azimuthally symmetric, it is likely that the variation of the radius of curvature of the emission in the two peaks is the cause of the differing hardness of their spectra (Barnard et al. 2018).  The highest-energy particles also dominate the ICS spectrum at VHE energies. 

The model light curve of the IR-optical emission consists of two peaks with a smaller phase separation than for the $\gamma$-ray peaks.  This morphology results from the restriction of this emission to regions of anti-GJ current and to lower altitudes, so that most observers will see the emission from only one pole, the one associated with the second $\gamma$-ray peak.  It is also the pole opposite to the one from which the radio peak is seen.  This geometry is very similar to that of the outer gap models, but in that case both $\gamma$-ray peaks originate from the magnetic pole opposite to that of the visible radio pulse.  RD17 used such an outer gap model to explain the IR-optical light curve.  In our model, the two $\gamma$-ray peaks originate  from different hemispheres, with the first peak in the same hemisphere as the visible radio peak.  The anti-GJ (return) current regions of the global magnetosphere thus define similar pair-active field lines to those of the outer gap model.  While the phases of the first and second peaks of the model IR light curve agree with those observed, their phases relative to the $\gamma$-ray peaks do not since, as mentioned above, our model $\gamma$-ray phases occur too late.

\section{Discussion}

We have modeled the emission from the Vela pulsar over a broad energy range from IR to 100 TeV with the primary goal of producing a spectrum of scattered IR radiation by the highest-energy particles.  From this study we obtained several major results.  We find that particles accelerated primarily in the current sheet of an oblique, near-force-free magnetosphere can produce a significant component of scattered IR/optical emission peaking around 20 TeV that is near the H.E.S.S. II sensitivity limit.  The model SR spectrum from electron-positron pairs produced in cascades near but on field lines inside the return current layer, and resonantly absorbing radio photons, very well matches the observed IR/optical spectrum in both amplitude and spectral index.  This spectrum extends to much lower energies, thus providing additional soft photons that the primaries can scatter in the Thompson regime.  This soft emission is located inside the light cylinder at relatively low altitudes of $0.2 - 0.5 R_{\rm LC}$, near and above the altitude of the radio emission.  The match of both the spectrum and the light curve thus locates the observed IR/Optical emission to these low altitudes.

Our model is operationally similar to that of RD17 but its physical foundation is fundamentally different.  While they use an outer gap model to accelerate primary particles between the null charge surface and the LC, we use a force-free geometry and, guided by results of global MHD and PIC models, accelerate particles primarily beyond the LC near the current sheet.  The primaries in our model produce a spectrum of scattered IR emission that is very similar to the spectrum obtained by RD17, who used a toy model of the IR/Optical emission distributed along the inner edge of the outer gaps.  We were able to show that the physical origin of the IR/Optical emission is SR from the pairs produced at the PCs in super-GJ regions and that the scattering of this emission from accelerating primary particles is similar to that produced by the toy models.  
RD17 include an extra ICS component from pairs scattering on thermal X-rays from the PC to give a peak in the IR light curve at the radio phase.

The detection of VHE emission from the Vela pulsar at energies above 3 TeV and possibly above 7 TeV requires the presence of particles accelerated to at least 10 TeV.  Particles having this energy in a pulsar magnetosphere will primarily radiate CR at GeV energies.  Furthermore, 10 - 30 TeV is the steady-state energy that the particles accelerating with the $E_\parallel$ constrained by the observed GeV emission will reach in the radiation-reaction limit when CR reaction is balanced by acceleration.  In contrast, pulsar models where particles are primarily accelerated by reconnection and radiate SR in the current sheet attain a maximum Lorentz factor of only $10^5 - 10^6$, or $\sim 0.1$ TeV (Petri 2012, Mochol \& Petri 2015, Cerutti et al. 2016, Philippov \& Spitkovsky 2018).  The observed VHE emission thus gives strong support to models where CR rather than SR is the dominant GeV pulsar emission mechanism.

\acknowledgments  
A.K.H. and C.K. acknowledge support from the National Science Foundation under Grant No. AST-1616632 and the NASA Astrophysics Theory Program.  C.V., A.K.H. and C. K. acknowledge support from the $Fermi$ Guest Investigator Program.  Resources supporting this work were provided by the NASA High-End Computing (HEC) Program through the NASA Center for Climate Simulation (NCCS) at Goddard Space Flight Center.   We thank Craig Pelissier of the NCCS in particular for help with parallel processing. 
This work is based on the research supported in part by the National Research Foundation
of South Africa (NRF; Grant Number 99072). The Grantholder acknowledges that opinions, findings and conclusions or recommendations expressed in any publication generated by the NRF-supported research is that of the author(s), and that the NRF accepts no liability whatsoever in this regard.

\newpage

\newpage


\end{document}